\documentstyle[12pt]{article}

\newcommand{\bmat}{\left(\begin{array}}
\newcommand{\emat}{\end{array}\right)}
\def\NPB#1#2#3{Nucl. Phys. B{#1} (19#2) #3}
\def\PLB#1#2#3{Phys. Lett. B{#1} (19#2) #3}

\def\PRD#1#2#3{Phys. Rev. D{#1} (19#2) #3}

\def\yzero{\smash{\hbox{$y\kern-4pt\raise1pt\hbox{${}^\circ$}$}}}

\def\-{\hphantom{-}}

\def\s2{\frac{1}{\sqrt2}}

\def\beq{\begin{equation}}
\def\eeq{\end{equation}}
\def\beqa{\begin{eqnarray}}
\def\eeqa{\end{eqnarray}}

\def\IF{\relax{\rm I\kern-.18em F}}
\def\II{\relax{\rm I\kern-.18em I}}
\def\IP{\relax{\rm I\kern-.18em P}}
\def\IC{\relax\hbox{\kern.25em$\inbar\kern-.3em{\rm C}$}}
\def\IR{\relax{\rm I\kern-.18em R}}

\def\Dsl{\,\raise.15ex\hbox{/}\mkern-13.5mu D} 
\def\IZ{Z\kern-.4em  Z}

\topmargin
0.5cm
\textwidth
15.5cm
\textheight
21.5cm
\oddsidemargin
0.7cm
\evensidemargin
1.2cm

\begin{document}

\makeatletter
\@addtoreset{equation}{section}
\makeatother
\renewcommand{\theequation}{\thesection.\arabic{equation}}
\pagestyle{empty}
\rightline{FTUAM-98/21; IFT-UAM/CSIC-98-26}
\rightline{McGill-98/33;  DAMTP-1998-147}
\rightline{\tt hep-ph/9810535}
\vspace{0.5cm}
\begin{center}
\LARGE{Strings at the Intermediate Scale \\
or \\
is the Fermi Scale Dual to the Planck Scale?\\[5mm] }
\large{C.P. Burgess${}^a$, L.E. Ib\'a\~nez${}^b$  and
F. Quevedo${}^c$ {\footnote{On leave of absence from Instituto de F\'{\i}sica,
UNAM, M\'exico.}}
\\[4mm]}
\small{
${}^a$ Physics Department, McGill University,\\[-0.3em]
3600 University St.,
Montr\'eal, Qu\'ebec, Canada, H3A 2T8.
\\[1mm]
${}^b$ Departamento de F\'{\i}sica Te\'orica C-XI
and Instituto de F\'{\i}sica Te\'orica  C-XVI,\\[-0.3em]
Universidad Aut\'onoma de Madrid,
Cantoblanco, 28049 Madrid, Spain.
\\[1mm]
${}^c$ D.A.M.T.P., Silver Street, Cambridge, CB3 9EW, England.\\[5mm]}
\small{\bf Abstract} \\[5mm]
\end{center}

\begin{center}
\begin{minipage}[h]{14.0cm}
We show that if the string scale is identifed with
the intermediate scale, $M_s=\sqrt{M_W M_{Planck}}
\sim 10^{11}$ GeV, then the notorious hierarchy,
$M_W/M_{Planck} \sim 10^{-16}$, can be explained
using only $M_c/M_s \sim 0.01 \sim  \alpha_{GUT} $
as small input parameters, where $M_c$ is the
compactification scale. This is possible for
weakly-coupled Type-I open-string vacua if the
observed world is assumed to live in an $N=1$ supersymmetric
3-brane sector coupled to a separate, hidden, 3-brane world
which breaks supersymmetry, because for such a model
$M_W/M_{Planck} = \frac12 \; \alpha_{GUT}^2 (M_c/M_s)^6$.
We discuss some of the phenomenological issues presented
by such an intermediate-scale string, showing that its benefits
include: ($i$) the possibility of logarithmic
gauge-coupling unification of the SM couplings
at $M_s$;
($ii$) a natural axionic solution to the strong-CP problem
with a phenomenologically-acceptable Peccei-Quinn scale;
($iii$) experimentally-interesting neutrino masses,
and more.

\end{minipage}
\end{center}
\newpage
\setcounter{page}{1}
\pagestyle{plain}
\renewcommand{\thefootnote}{\arabic{footnote}}
\setcounter{footnote}{0}

\section{Introduction}

It has recently been realized that the traditional
connection between the string scale and the Planck mass,
$M_s= \frac12 \; \sqrt{\alpha _{GUT}} M_{Planck}$ (as
is found in perturbative heterotic string theory) need
not apply to all string vacua \cite{witten,lykken,
untev,breathmode,bajogut,shiutye,bachas,kakutye,
benakli,bendav}.

In this paper our purpose is to argue that the string scale
is the geometric mean of the weak and Planck scales:
$M_s = \sqrt{M_W M_{Planck}} \sim 10^{11}$ GeV.
Our main argument in favour
of this proposition is that its adoption permits the
problematic hierarchy, $M_W/M_{Planck} \sim 10^{-16}$, to
be completely understood without using any dimensionless inputs
which are smaller than 1\%. We believe this represents considerable
improvement over other explanations of the hierarchy
problem, which typically explain the small value of
$M_W/M_{Planck}$ in terms of another kind of hierarchy.
For instance, the recently-proposed intriguing possibility
that $M_s$ might be as low as the TeV scale \cite{lykken,untev,
breathmode,bajogut,shiutye,bachas,kakutye,benakli,bendav},
requires a compactification scale, $M_c$, which satisfies
$M_c/M_s \le 10^{-5}$.

Besides ameliorating some of the phenomenological
problems of the TeV-scale scenario (such as too-fast
proton decay), we find a number of other good
features follow if $M_s$ is identified
with $10^{11}$ GeV. These include the possibility of logarithmic
perturbative unification (at $M_s$) of the Standard Model
(SM) gauge couplings; naturally-occurring axions having
astrophysically-acceptable couplings and Peccei-Quinn (PQ)
scales; potentially interesting neutrino masses, etc.

Our arguments are based on the more general connection between
$M_s$ and $M_{Planck}$ which occurs in perturbative Type-I
string theory:
\beq
{ {M_c^{(p-6)}} \over {M_s^{(p-7)} } } \ =\
{ {\alpha _p M_{Planck} }\over {\sqrt {2} }  }
 \
\label{const}
\eeq
where $p$ corresponds to the apropriate p-brane from
which the gauge group (with coupling $\alpha _p$)
originates. In addition, the condition that we remain within
the realm of perturbation theory requires
the corresponding $D=10$ dilaton coupling, $\lambda_s$, to obey:
\beq
\lambda_s\ =\ 2\alpha _p \ \left({ { M_s}\over {M_c} }
\right)^{p-3} \ = \
2\sqrt{2} { { M_s^4}\over {M_c^3M_{Planck} } }\  \leq  O(1).
\label{pert}
\eeq
These two conditions require (in this simple
isotropic case) the compactification scale, $M_c$,
to not be much smaller than the string scale. Apart
from this there is a remarkable freedom in choosing
these scales.

There are, however, several natural options for these scales
which suggest themselves. In order to explore these,
consider for definiteness an embedding of the
SM interactions within a set of coincident 3-branes.

\begin{enumerate}
\item
The first possibility is to conform with perturbative
heterotic tradition, and to place the string scale not
far from the Planck scale.
\item
A slightly better idea \cite{witten} is to identify
$M_s$ with the GUT scale, $M_X=M_s=2\times 10^{16}$ GeV,
as indicated by the extrapolation of the low-energy
couplings in the MSSM. This is consistent with
the value of $M_{Planck}$ inferred from (\ref{const}), if we
appropriately choose $M_c$ very slightly below $M_s$.
Thus, in this way the old problem of reconciling
gauge-coupling unification with perturbative heterotic
strings is naturally solved. On the other hand, this
scenario offers no explanation for the origin of the
huge hierarchy of scales between the Fermi scale, $M_W$,
and $M_{Planck}$, which must instead be blamed on some
non-perturbative mechanism, like gaugino condensation.
\item
Another alternative which has received recently much
attention \cite{lykken,untev,breathmode,bajogut,shiutye,
bachas,kakutye,benakli,bendav} is the possibility of
bringing $M_s$ down to the weak scale: $M_s\propto 1$ TeV.
This is a very intriguing possibility, raising as it does
the possibility of testing string theory at accelerators.
In this case, the choice $M_c/M_s\le 10^{-5}$ is
required in order to obtain the correct value of $M_{Planck}$.
One trades in this way the standard hierarchy,
$M_W/M_{Planck}$, for this less extreme, but small, ratio.
On the other hand, reconciling this scheme with
constraints from cosmology, proton stability and
gauge-coupling unification may prove non-trivial.
\end{enumerate}

As stated above, we here propose to identify the string scale
with the intermediate scale,
$M_s=\sqrt{M_WM_{Planck}}\propto 10^{11}$ GeV. This scale
arises in a number of phenomenological settings,
as we discuss below. The most interesting point of
this scenario is the economy of its explanation
of the $M_W/M_{Planck}$ hierachy, which here arises
from the amplification of an initially very modest
supression, $M_c/M_s\propto 10^{-2}$. This
amplification occurs, without the need for any special
hierarchy-generating mechanism such as gaugino condensation,
due to the large powers of $M_c/M_s$ which appears
in eq.~(\ref{const}).

(After completing this paper we discovered some earlier
work by Benakli \cite{benakli} (see also \cite{bachas})
which explores some of the advantages (including the
connection with invisible axions and neutrino masses) of
a having the string scale of order $10^{11}-10^{14}$ GeV.
Different scenarios in possibilities for the
generation of SUSY-breaking in the brane context have
been also considered in refs.~\cite{kakutye,untev,benakli,randrum}.
Similar studies in the context of the Horawa-Witten
M-theory scheme can be found in refs.\cite{mteoria} .)

To see how the $M_W/M_{Planck}$ hierarchy arises, we must
determine how $M_W$ depends on the basic scales, $M_c$ and $M_s$.
This dependence arises once supersymmetry breaks, which take
to happen in a hidden sector of the model.
Hidden sectors arise naturally within Type I string vacua.
Let us consider for definiteness two separate sets of
parallel 3-branes. We imagine ourselves to live
on one set, containing the SM, while the other
contains the hidden-sector interactions.
If the positions of these 3-branes in the transverse space
are sufficiently different, there are no massless states charged
under both 3-brane groups. In what follows we consider
an ideal situation of this type, in which we have
the set of 3-branes containing the SM particles have
unbroken $N=1$ SUSY, and the distant (hidden) set of
parallel 3-branes somehow completely break SUSY.
We need not make any particular assumption about
the nature of the SUSY breaking in that sector.

SM particles do not directly feel the breaking of
SUSY which takes place in the hidden 3-brane sector.
The effects for the SM of SUSY-breaking are only
transmitted through the influence of closed-string
sector fields, which are the only ones which can move
into the bulk of spacetime and couple to both kinds of 3-brane
sectors. Thus the SUSY-breaking felt by the
SM fields are automatically supressed by powers of
$M_{Planck}$. In particular, if no particular
supression of SUSY-breaking in the hidden sector
is assumed, one expects that the SUSY-breaking soft terms
felt in the SM sector are of order:
\beq
 M_W \ \ \sim m_{3/2} \ \sim \ {{F}\over {M_{Planck} } }
\ \sim \ { { M_{s}^2}\over {M_{Planck}} }
\label{debil}
\eeq

Consider next, for simplicity, that all six of the compact
dimensions share a common overall compactification
scale, $M_c$. From the above formulae one has:
\beq
{ {M_W}\over {M_{Planck} }  }\ =\
 { {\alpha _3^2}\over {2} }\ \left({ {M_c}\over {M_s} }
\right)^{6}
\label{jerar3}
\eeq
Or, more specifically:
\beq
M_W\ =\ {{\alpha _3}\over {\sqrt{2}} } { {M_c^3}\over {M_s^2} }
\ ;\ M_{Planck}\ =\ {{\sqrt{2}}\over {\alpha _3} }
{ {M_s^4}\over {M_c^3} } .
\label{mwmp}
\eeq
Notice that SUSY-breaking dissapears as $M_c\rightarrow
0$ and the distance between visible and hidden
3-branes go to infinity. Furthermore,
if one takes $M_s/M_c\propto 160 $ and $\alpha _3=1/24$ one
indeed obtains the desired hierarchy $M_W/M_{Planck}
=10^{-16}$. As claimed, this small ratio
arises from the large power of $M_c/M_s$ which
amplifies a modest input value for $M_c/M_s$.
It is remarkable how such a large hierarchy of
16 orders of magnitude can so naturally appear
from such a modest initial supression
$M_c/M_s\propto 10^{-2}$. We regard this initial
`hierarchy' of $10^{-2}$ to be no hierarchy at all,
since such small numbers are easy to obtain.
Furthermore, we know that small numbers of this order
have to appear elsewhere in the theory anyhow,
such as if we are to understand the
intergeneration ratio of Yukawa couplings.

Several remarks are worth recording before passing on
to the phenomenological implications of this scenario.

\begin{enumerate}
\item
A similar argument works equally well for the $p=9$ case,
in which similar formulae are obtained, but with the
replacements $\alpha _3\leftrightarrow \alpha_9$ and
$M_s/M_c\leftrightarrow M_c/M_s$. In this case the
required input factor would be the inverse of that just discussed,
$M_c/M_s \propto 160$. The condition that we remain within
Type I perturbation theory further requires $\lambda_s
= 2\alpha_p(M_s/M_c)^{(p-3)} <1$. This is satisfied for
both 3-branes and 9-branes. In the 9-brane case, however
the Type I dilaton coupling $\lambda_s$ would have to be very
small and the scheme becomes less natural.
\item
Next, we remark in passing on another fascinating connection between
the two derived scales, $M_W$ and $M_{Planck}$, which are
generated from the two fundamental scales of the model,
$M_s$ and $M_c$. If we write $\alpha' \sim 1/M_s^2$ then
$M_W$ and $M_{Planck}$ are related one to the other
by $M_W= 1/(\alpha 'M_{Planck})$. This is reminiscent
of a `T-duality' relationship, raising the tantalizing
speculation that perhaps in some sense the physics
at the Fermi scale might turn out to be `T-dual'
to the physics at the Planck scale.
\item
Finally, we ask whether a similar explanation of the
hierarchy problem is possible within the
Horava-Witten representation of strongly coupling heterotic
strings. We find the situation to be different in this
case, for which the following relations obtain \cite{witten} :
\beq
{M_{Planck}^2}\, =\, \frac{M_m^9}{M_c^6 M_{\rho}}
\qquad \alpha_9=(\sqrt{2}\pi)^{2/3} \left(\frac{M_c}{M_m}
\right)^6 ,
\eeq
where $M_m$ is the 11-dimensional $M$-theory mass scale, $M_{\rho}$ is
the mass scale associated to the 1-dimensional interval and the other
parameters are like in type-I theory. If we have that
$M_W=M_m^2/M_{Planck}$
then we obtain the following relations:
\beq
\frac{M_W}{M_{Planck}}= \left(\frac{M_c}{M_m}
\right)^6 \; \left(\frac{M_\rho}{M_m} \right)
\qquad
\alpha_9= (\sqrt{2}\pi)^{2/3} \left(\frac{M_c}{M_m}
\right)^6
\eeq
Therefore, from the first relation it seems that we can also obtain the
hierarchy by a small hierarchy between the compactification scales and
the string scale. However, the second relation tells us that if the
compactification scale $M_c$ is very small compared with the M-theory scale,
then the gauge coupling would be far too small to agree with
experiment. We can still use a standard value for
$\alpha_9$ and obtain the hierarchy between the electroweak and Planck
scales, but only in terms of a similarly large hierarchy
between the M-theory scale and the length of the
11-dimensional interval. The ultimate origin of this
hierarchy then remains unexplained.
On the other hand, contrary to the case with weak-scale M-theory,
it is in principle possible to have an intermediate scale
M-theory without having an unacceptably large value for the
interval length, $\rho$, since the observed hierarchy
implies in this case $\rho \sim 10^{-12}$ meters.

\end{enumerate}

\section{Phenomenological Issues}

A number of questions and phenomenological implications
appear in this scheme:

{\bf i) Gauge coupling unification}

If the string scale is of order $10^{11}$ GeV, one would also expect
that gauge couplings of the SM should also join at this scale.
We now argue that this kind of unification does not need
fast power-like running, as would be mandatory
for a weak-scale-string scenario. Indeed, if there are
further particles charged only under $SU(2)\times U(1)$ with masses
of order $M_W$ in the massless spectrum beyond those of the MSSM,
the $SU(2)\times U(1)$ couplings would grow faster and so intersect
the $SU(3)$ coupling precociously.

To see this, recall
the one-loop formulae for the running of the SM gauge couplings:
\beqa
sin^2\theta _W(M_Z)\ =&\ {3\over 8}(1\ +\ {{5\alpha (M_Z)}\over
{6\pi }}\ (b_2-{3\over 5}b_1)\ log({{M_s}\over {M_Z}})\ )  \\ \nonumber
{1\over {\alpha _3(M_Z)}}\ =&\ {3\over 8}({1\over {\alpha (M_Z)}}\ -\
{1\over {2\pi }}\ (b_1+b_2-{8\over 3}b_3)\ log({{M_s}\over {M_Z}})\ )  \\
\label{senos}
\eeqa
where $b_i$ are the SM beta-function coefficients. In the MSSM one has
$b_1=11, b_2=1, b_3=-3$ yielding unification at $2\times 10^{16}$ GeV.
In the present case one can check that increasing
$b_2$ by three units makes $\alpha _2$ cross
$\alpha _3$ at around $M_s=10^{11}$ GeV as required.
In order to obtain consistent values with
both $\alpha _3$ and $sin^2\theta _W$ one has further to increase
the value $b_1$ by around eleven units. An example of additional
particles which can produce these beta functions is
given by supplementing  the three SM quark-lepton generations with
a collection of new chiral fields transforming like e.g.
\beq
4[ (1,1,1)+(1,1,-1)+(1,2,1/2)+(1,2,-1/2)]  \ .
\eeq
under $SU(3)\times SU(2)\times U(1)$. These are just
 four standard sets of of vector-like leptons. A pair of
doublets in this case should be identified with the SM Higgs.
 In this case one
finds $\alpha _3(M_Z)=0.12$ and $sin^2\theta _W(M_Z)=0.23$
for $M_s=3\times 10^8 M_W$. Of course this particularly
simple choice is not unique, and other combinations
could be possible.  As discussed below,
such an extension of the MSSM might be interesting if one wants to
gauge a symmetry like lepton number to insure proton stability.

Another alternative is to consider extensions to the SM
gauge group like $SU(4)\times SU(2)_L\times SU(2)_R$. (Such
a group often appears in orientifold constructions.)
This kind of gauge group can also lead to precocius unification.
Clearly, a more systematic study of the unification
possibilities at the geometrical scale $M_s\propto 10^{11}$ GeV
would certainly be interesting.

We conclude that gauge coupling unification can be easily
accomodated in this scheme, although it is not,
properly speaking, predicted.

{\bf ii) Proton stability}

It is well known that generic versions of the SUSY SM contain
$D=4$ operators wich violate baryon- and/or lepton number unless
some symmetry (like e.g., the standard discrete R-parity)
is imposed on the model. In addition, even if those
$D=4$ terms are forbidden,
if the string and unification scales are of
order of the intermediate scale
in general there will appear dimension 5 and
dimension 6 (and even higher)
operators which will violate baryon and lepton
numbers.

Thus there must exist
a symmetry forbiding these unwanted effects
also. The problem of dimension five operators
(although in a less severe form ) is in fact already present
\cite{pdecay}  in the
scheme with $M_s=10^{16}$ GeV  and is much more problematic in the
schemes with $M_s=1$ TeV. Thus this is really a problem common to
all schemes with $M_s<M_{Planck}$.
A  natural solution is the presence of
 a continuous gauged $U(1)$ symmetry.
In this context a symmetry like  $U(1)_{B-L}$ which appears in left-right
symmetric models  may forbid R-parity violating operators of dimension 4
but again is not going to forbid some dangerous dimension 5 operators.

Another natural alternative is
to consider a pseudoanomalous $U(1)_X$  whose
triangle anomalies are cancelled via the 4-dimensional Green-Schwarz
mechanism. The simplest flavour independent
such a $U(1)_X$ one can think of is
one which asigns charge=1 for all SM quarks and leptons,  charge= -2 for
the Higgs doublets $H,{\bar H}$. This simple symmetry naturally
forbids baryon and lepton violating dimension-4 and 5 operators.
Alas, in the case of perturbative heterotic vacua
this mechanism is very restrictive. Indeed for the mechanism to be
possible the mixed anomalies $A_i$  of this $U(1)_X$ with the SM  gauge groups
$G_i$, i=1,2,3 have to satisfy  $A_1:A_2:A_3=5/3:1:1$
\cite{ibanez} . In the case of
the MSSM one finds instead:
\beq
 A_1:A_2:A_3\ =\ 6:4:6  \ .
\eeq
 If e.g., one adds the
contribution of the extra states needed for gauge coupling unification
which we suggested in the previous paragraph on has more freedom.

  But in fact everything is simpler in the
class  of Type I  $D=4$ strings
that we are considering. It has been recently realized \cite{iru}  that
in the context of Type IIB $D=4$ orientifolds there is a generalized
Green-Schwarz mechanism at work in which the mixed anomalies
{\it are not}  constrained to be in definite ratios. This is because
of the generic presence of several twisted RR axionic fields
which couple differently to different group factors. Thus an
anomalous $U(1)_X$ like the one proposed above can be
perfectly consistent within the context of Type I theories
without the need of choosing particular charges for the
extra particles present to get gauge coupling unification.

A $U(1)_X$ symmetry with quark and lepton charges as considered
here was first considered by Weinberg  \cite{weinberg}
in the early days of SUSY model-building
in order to avoid proton decay via dimensio-4
and -5 operators. However he introduced extra
exotic particles to get cancellation of $U(1)$ anomalies.
This is not required in the context of string theory.
 In any event the
presence of such an  anomalous
$U(1)$
 is enough to forbid $D=4,5$ operators.
This is true even if the $U(1)$ becomes massive
by swallowing RR axionic field, because then the symmetry would
survive perturbatively as a global $U(1)$ symmetry.

As explained above, a simple anomalous $U(1)$ like the one discussed above
only supresses dimension 4 and 5 baryon/lepton number violating operators,
but not dimension 6  which need also to be supressed if the
string scale is as low as $10^{11}$ GeV. However, the relevant operators
will always involve quarks and lepton of the first
two generations and
hence one expects further supression factors.
A classical example for this are the dimension 6
operators coming from the exchange of colour-triplet
Higgs fields in the minimal SUSY-$SU(5)$ operator. From the
proton-stability
 point of view the
Higgs triplet can be as light as
$10^{11}$ GeV because the triplet couples
with very supressed Yukawa couplings
to the first two generations.

Let us finally remark that anomalous $U(1)$s with similar general
characteristics
to the one we are proposing here do appear
in specific $D=4$, $N=4$
Type IIB orientifolds (see ref.\cite{shiutye,iru}).
In particular, $SU(n)$
groups come along with $U(1)$ factors with the
same couplings in $U(n)$
gauge groups. Thus, e.g., if we have a set of 3-branes
with the SM non-Abelian gauge group $SU(3)\times SU(2)$ one indeed
expects to have $U(3)\times U(2)$ factors. One linear combination
of the two $U(1)$s would be the hypercharge whereas the orthogonal
one would typically  be  an anomalous $U(1)$.

Another alternative is the presence of anomaly-free
$U(1)$s gauging either lepton or baryon numbers. As
explained in \cite{ibaross}   this is possible if one adds
vector-like sets of leptons to the MSSM.
We have seen that extra vector-like
sets of leptons are wellcome in order to
get gauge coupling unification,
thus this is  an alternative which could be at work.

{\bf iii) Soft terms and radiative electroweak symmetry breaking}

In a simple scheme in which we assume that the SM  and the
hidden sector live in two separated sets of p-branes, only
closed-string fields are able to transmit SUSY-breaking
from the hidden to the visible sector. Thus it is natural to assume that
{\it the mediators of SUSY-breaking will be the dilaton and
moduli fields}. Specifically, in the context of Type IIB
$D=4$, $N=1$ orientifolds the complex dilaton $S$ and
untwisted moduli fields $T_i$ are able to couple to both
sectors of p-branes. This is not in general the case
for twisted closed-string moduli,  which couple only
to sets of p-branes with positions in transverse space
on top of the given orbifold singularities. In any event,
it would make sense from the visible sector point of view
to parametrize SUSY breaking in terms of the vaccum
expectation values of the auxiliary fields of
dilaton and untwisted moduli, $F_S, F_{T_i}$.
This is the spirit previously applied to heterotic compactifications
in refs. \cite{spurion,bim,bims}
 and generalized to Type I  type of vacua in ref.~\cite{imr}.
Thus possibilities like dilaton/modulus dominated boundary
conditions may be particularly relevant in the present scheme.
However one does not need to assume that $F_S$ and $F_{T_i}$
are the only fields contributing to the vacuum energy since
in the hidden sector there may be other fields (e.g.
the  twisted
moduli in Type IIB orientifolds) who contribute
 to SUSY-breaking but which do not couple
directly to the visible p-branes.

The existence of fields which can contribute to SUSY-breaking but
which do not couple to the visible p-brane sector somewhat
changes  the results for soft terms from dilaton/moduli
dominance as computed in heterotic
models. As explained in \cite{afiv,imr},
the structure and couplings
of the massless p-brane sector in Abelian Type IIB $D=4$
orientifolds is quite analogous to the untwisted sector of
Abelian heterotic orbifolds and so is the Kahler
potential (modulo some redefinitions of the $S$ and $T_i$
chiral fields) and renormalizable superpotential. Thus
one obtains similar soft terms results as those
found for the untwisted sector of heterotic orbifolds.
The case of the existence of some field $\phi $ contributing to
SUSY-breaking but not coupling to the visible world was
in fact considered in chapter 8 of \cite{bim}.  One finds
for the case of an overall modulus field $T$ the result for
the soft scalar and gaugino masses (assuming a
vanishing cosmological constant):
\beq
m_0^2\ = m_{3/2}^2 ( 1-cos^2\theta cos^2\theta _{\phi } )
\ ;\
M_{1/2}^2\ =\ 3m_{3/2}^2 sin^2\theta
\label{soft}
\eeq
where $tg\theta =|F_S|/(\sqrt{3}|F_T|)$ and $sin\theta _{\phi }
 =|F_M|/(\sqrt{3}m_{3/2})$.
Thus in the absence of a SUSY-breaking contribution from $\phi $ one has
$cos\theta _{\phi }=1$ and one goes back to the usual dilaton/modulus dominated
limit with $M^2_{1/2}=3m_0^2$.
On the other hand if there are fields $\phi $
contributing to SUSY-breaking but not coupling to the visible p-branes
(like e.g. the twisted fields we mentioned above) one sees that
smaller gaugino masses with $M_{1/2}\leq \sqrt{3}m_0$ are now possible.
Notice that  in the above expressions one has
$m_{3/2}\propto \alpha _3/2(M_c/M_s)^6$.

Unlike the situation in the weak-scale string
scenarios, in the present scheme,
 once  SUSY breaking of order
$\alpha_3/\sqrt{2}(M_c^3/M_s^2)$ is transmitted to
the visible 3-brane sector, radiative $SU(2)\times U(1)$
breaking occurs in the usual way, since there is
plenty of room for the evolution of
the soft masses from $10^{11}$ GeV to the weak scale.
Since, nevertheless, the space for running is
substantially smaller than in the usual MSSM scheme,
for a given set of soft terms, radiative breaking in the present scheme
will in general require higher top-quark Yukawa couplings.
This is because larger top-quark Yukawa couplings make
the Higgs field squared-mass run faster towards smaller values.

{\bf iv) Invisible axions and the Strong CP problem}

One of the bonuses of the present scheme is that it seems to
provide a general scenario for the solution of the strong CP problem.
It is well known that astrophysical constraints
\cite{redgiants} coming
from the stability of red-giants impose the lower bound on the
decay constant of an axion $M_{PQ}>10^{10}$ GeV. In fact there
are also cosmological bounds which give an upper bound
of the same order of magnitude, fixing the Peccei-Quinn scale
around the intermediate scale.
In any event, it is well known
that in Type I string theory there may be a plethora of RR scalars
with axion-like couplings (like the
twisted RR fields  mentioned in the previous subsection).
The typical scale of the corresponding decay constants
is expected to be of order $M_s=10^{11}$ GeV,
consistent with the astrophysical bounds.
Thus the RR scalars of Type I theory provide naturally
with the required invisible axions at the apropriate
mass scale.

String axions appearing in perturbative heterotic string theory
have been studied in the past as possible candidates for
invisible axions \cite{stringaxions}.
In particular the model-independent
ImS axion as well as the  imaginary partners of $T$ moduli which
can also get axionic couplings at one loop. However in the
perturbative heterotic case the natural Peccei-Quinn scale
is the string scale or slightly below and thus it is too large.
Also it is difficult to avoid that these would be axions
would get a mass upon SUSY-breaking from space-time or
world-sheet instantons.

In the present scheme the string scale coincides with the
Peccei-Quinn scale, so this  part of the problem is solved.
On the other hand in e.g., Type IIB orientifolds
 not only the untwisted moduli
$S,T_i$ but also twisted Ramond-Ramond (RR) fields have
axionic couplings
\cite{iru} . Although $ImS$ and or $ImT_i$
might get too-large masses from interactions with the
$N=0$,  3-brane sector, that is not going to be in general the
case for some  twisted RR-fields $ImM_a$. Indeed, these fields
couple only to 3-branes situated on the corresponding
orbifold singularity (whose blowing-up mode is related to
$ReM_a$). Thus {\it $ImM_a$ axionic fields coupling to SM gauge
groups will not couple to hidden sector groups}. This will
guarantee that only QCD instantons will determine the
potential of these axions. This will then lead to
an automatic solution of the strong CP problem.

{\bf v) Neutrino masses}

Intermediate scales of order $10^{10-13}$ GeV are popular
in the literature in the context of see-saw models
of neutrino masses. Integrating out any heavy Majorana neutrinos,
whose masses we suppose are of order $M_s$, generates the
effective interaction $LLHH$ in the low-energy
superpotential, where $L$ and $H$ represent SM lepton
and higgs doublets. The coefficient of such an operator is
of order $a/M_s$, where $a$ contains products of neutrino
Yukawa couplings. The Majorana masses which follow for
the light neutrino species are then $m_\nu \sim a
\langle H \rangle^2/M_s \sim a (10 \; \hbox{eV})$.

For reasonable values for $a$, masses this size can
lie below the experimental limit (from $\beta\beta$
decay) of $\sim 1$ eV for the electron neutrino, and
can easily lie in a range which is consistent with
atmospheric, solar and/or LSND results. (Accomodating
all three typically would require light sterile neutrinos,
which are also possible in the scenario we are considering.)

Thus in the present scheme the right-handed
neutrinos could be just massive string states with masses of order
$M_s$, making current neutrino experiments windows
onto string physics!

{\bf vi)  Wimpzillas and ultra-high energy cosmic rays}

Intermediate scales have also recently appeared in the context of
supermassive, stable,  particles with masses of
order $10^{12}-10^{16}$ GeV
(the so-called `Wimpzillas' \cite{wimpzillas}) which
could constitute an interesting candidate
for cold dark matter. Particles associated to the string scale
could provide candidates for such states. Furthermore,
decaying particles asociated to the same string scale could be candidate
 sources to generate ultra-high energy cosmic rays \cite{steve}\
(see \cite{cosmic} and references therein) with
energies $\propto 10^9-10^{10}$ GeV found experimentally.

{\bf vii) Cosmology }

The cosmological implications of the intermediate scale may be many.
Although usually inflation is considered much closer to the  Planck
scale, there is no impediment for it to occur at an intermediate
scale. The amount of fine tuning needed to obtain the expected number
of $e$-foldings and the spectrum of perturbations is
of course model-dependent.
For a single inflaton field this will imply an inflaton mass of the
order of $M_s^2/M_{Planck}$, {\it ie} the electroweak mass, which is
quite reasonable, compared
with a $10^{-13}$ GeV inflaton in the TeV string scenario \cite{bendav} .
Furthermore with an intermediate scale string theory it is
straightforward to obtain large enough reheating temperatures in order
to produce the standard model fields after inflation \cite{bendav}. Being
sufficiently higher than the electroweak scale, there should not be any
problem to generate the baryon asymmetry. A detailed  study of all these
issues may be interesting.

{\bf viii) Generating $M_c$ and the breathing mode}

Considerable effort \cite{breathmode} has been devoted to
understanding the hierarchy $M_c/M_s$ in models
with a weak-scale string scale. This proves to be reasonably
difficult to do, due to the partially-conflicting constraints
that: ($i$) $M_c$ be small enough to give the correct
value for $M_{Planck}$, and yet that the breathing mode,
$T$, of the compact dimensions not be so light as to be
ruled out by constraints on long-range, gravitational-strength
scalar interactions.

Clearly a hierarchy like $M_c/M_s \sim
10^{-2}$ is much easier to generate, to the extent
that it does not require much explanation at all. Furthermore
the constraints on the mass of the breathing mode in
our scenario are also much weaker. This is because
the mass of any such mode can be at most as large
as $M_c$. Since for energies larger than
$M_c$ the breathing mode is a component of the
extra-dimensional metric, and so is required to
be massless by general covariance. Generically the
breathing mode can be much lighter than this upper
bound. Direct experimental limits on `fifth forces'
preclude this mass being smaller than an inverse millimetre,
which is easy to satisfy when $M_c \sim 10^9$ GeV, but
more difficult when $M_c \ll M_s \sim 1$ TeV.

\section{Conclusions}

In summary, we have proposed as a natural  scheme one in which
the string scale $M_s$ is identified with the intermediate scale
$M_s=\sqrt{M_WM_{Planck}}  $. An important property of this scenario
is that the hierarchy $M_W/M_{Planck}$ appears as a consequence
of the compactification scale being just a couple of orders of magnitude
below the string scale, due to the large power appearing in
$M_W/M_{Planck}=\frac12 \; \alpha _3^2 (M_c/M_s)^6$.
This natural generation of
the huge hierarchy is a nice feature compared to the standard
assumption that both string and Planck scales are
close to one another.
Essentially what we
assume is the existence of a perturbative Type I string vacuum
with an $N=1$ supersymmetric 3-brane world (including the SM)
and a separate 3-brane world with no supersymmetry.

We remark that the weak and Planck scales in this scheme
are related by a `dual-looking' formula: $M_W=1/\alpha '
M_{Planck}$. It would be interesting to see whether this
is a reflection of some underlying duality between the
physics of these two scales.

Our scheme can easily accomodate logarithmic gauge-coupling unification
at the price of supplementing the particle content of the MSSM
just above the weak scale.
This is not as natural as in a conventional grand-unified scenario, but
is certainly simpler than gauge-coupling  unification in
the 1 TeV scenarios, which requires power-like running of
coupling constants.

Another attractive aspect of the present
scheme is that the string scale turns out to coincide with the
Peccei-Quinn scale that is required by astrophysical constraints
for invisible axion models which solve the strong
CP problem. Since string theory has abundant
axion-like fields (like e.g., the twisted RR fields in Type
I orientifold models), and some of these may couple to the
SM but not to the hidden sector,
this problem could be naturally solved.

String theory at an intermediate scale could also be relevant
for other phenomenological issues like neutrino masses
and cosmology.
In any event, it seems that if indeed the string scale is of order
the intermediate scale $\sim 10^{11}$ GeV, string theory
could be much more amenable to experimental test than previously thought.

\bigskip

\bigskip

\bigskip

\centerline{\bf Acknowledgements}
We acknowledge useful discussions with C. Mu\~noz, R. Myers and
N. Turok. F.Q. acknowledges  PPARC for an advanced research
fellowship. During the development  of this research project he was a
 John Simon Guggenheim fellow.

\end{document}